\documentclass[bibyear]{aa} 

\usepackage{txfonts}
\usepackage{graphicx}   
\usepackage{amsmath}    
\usepackage[utf8]{inputenc} 
\usepackage{array} 
\usepackage{multicol,rotating} 
\usepackage{longtable} 
\usepackage[section]{placeins}
\usepackage{lscape}

\begin{document}

   \title{Small lobe of comet 67P: Characterization of the Wosret region with ROSETTA-OSIRIS}

   \author{S. Fornasier
          \inst{1,2}
          \and
          J. Bourdelle de Micas\inst{1} \and
          P.H. Hasselmann\inst{1} \and
          H. V. Hoang\inst{1,3} \and
          M.A. Barucci\inst{1} \and
          H. Sierks\inst{4}
          }

   \institute{LESIA, Observatoire de Paris, Universit\'e PSL, CNRS, Universit\'e de Paris, Sorbonne Universit\'e, 5 place Jules Janssen, 92195 Meudon, France
              \email{sonia.fornasier@obspm.fr}
         \and
           Institut Universitaire de France (IUF), 1 rue Descartes, 75231 PARIS CEDEX 05  
        \and  
        Universit\'e Grenoble Alpes, CNRS, Institut de Plan\'etologie et Astrophysique de Grenoble (IPAG), UMR 5274, Grenoble F-38041, France
        \and
          Max-Planck-Institut f\"ur Sonnensystemforschung, Justus-von-Liebig-Weg, 3, 37077, G\"ottingen, Germany
             }

   \date{Received on April 2021 }

 
  \abstract
   {}
   {We investigated Wosret, a region located on the small lobe of the 67P/Churyumov-Gerasimenko comet subject to strong heating during the perihelion passage. This region includes the two last landing sites of the Philae lander as well as, notably the final one, Abydos, where the lander performed most of its measurements. We study  Wosret in order to constrain its compositional properties and its surface evolution. By comparing them with those of other regions, we aim to identify possible differences among the two lobes of the comet.}
   {We analyzed high-resolution images of the Wosret region acquired between 2015 and 2016 by the Optical, Spectroscopic, and Infrared Remote Imaging System (OSIRIS) on board the Rosetta spacecraft, at a resolution ranging from 2-10 m/px before and close to perihelion, up to 0.07-0.2 m/px in the post-perihelion images. The OSIRIS images were processed with the OSIRIS standard pipeline, then converted into $I/F$ radiance factors and corrected for the viewing and illumination conditions at each pixel using the Lommel-Seeliger disk function. Spectral slopes were computed in the 535-882 nm range.}
   {We observed a few morphological changes in Wosret, related to local dust coating removal with an estimated depth of $\sim$ 1 m, along with the formation of a cavity measuring 30 m in length and 6.5 m in depth, for a total estimated mass loss of 1.2 $\times$ 10$^6$ kg. The spectrophotometry of the region is typical of medium-red regions of comet 67P, with spectral slope values of 15-16 \%/(100 nm) in pre-perihelion data acquired at phase angle 60$^o$. As observed globally for the comet, also Wosret shows spectral slope variations during the orbit linked to the seasonal cycle of water, with colors getting relatively bluer at perihelion. Wosret has a spectral phase reddening of 0.0546 $\times 10^{-4}$ nm$^{-1} deg^{-1}$, which is about a factor of 2 lower than what was determined for the nucleus northern hemisphere regions, possibly indicating a reduced surface micro-roughness due to the lack of widespread dust coating. A few tiny bright spots are observed and we estimate a local water-ice enrichment up to 60\% in one of them. Morphological features such as "goosebumps" or clods are widely present and  larger in size than similar features located  in the big lobe.}
   {Compared to Anhur and Khonsu, two southern hemisphere regions in the big lobe which have been observed under similar conditions and also exposed to high insolation during perihelion, Wosret exhibits fewer exposed volatiles and less morphological variations due to activity events. Considering that the high erosion rate in Wosret unveils part of the inner layers of the small lobe, our analysis indicates that the small lobe has different physical and mechanical properties than the big one and a lower volatile content, at least in its uppermost layers.  These results support the hypothesis that comet 67P originated from the merging of two distinct bodies in the early Solar System.
}


   \maketitle
%

\section{Introduction}

\begin{figure*}
\includegraphics[width=0.99\textwidth,angle=0]{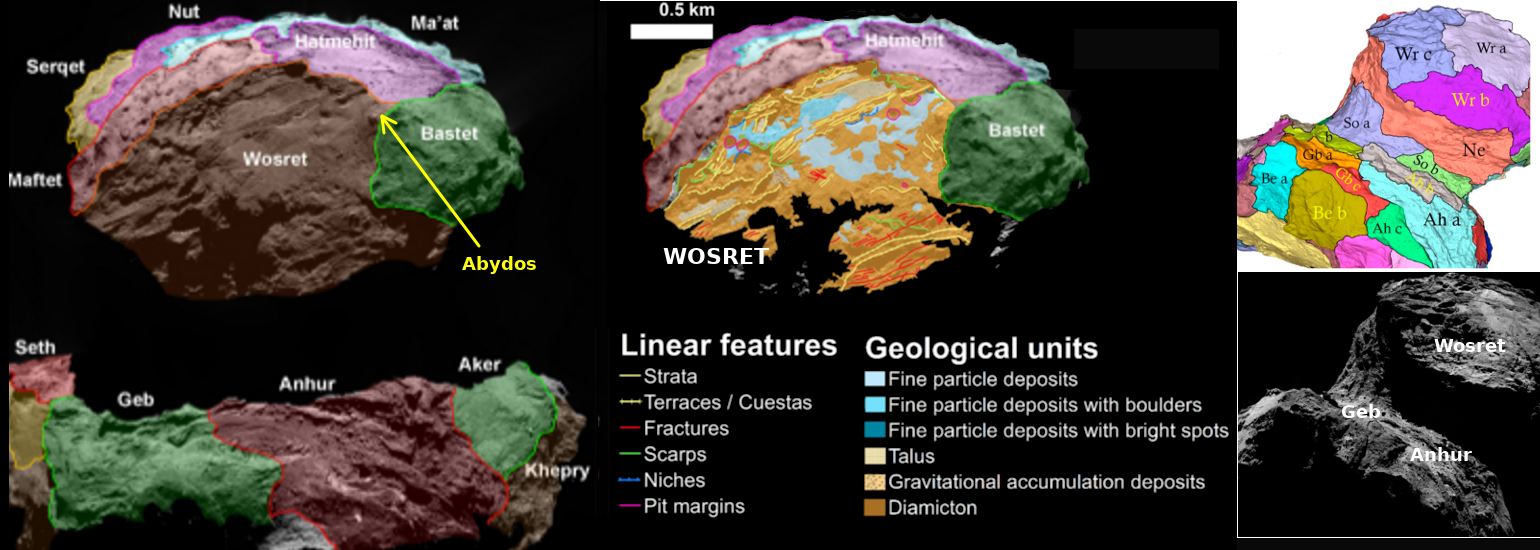}
\caption{The Wosret region on comet 67P. Left: 3D view of the southern hemisphere of 67P (from an image acquired on 2 May 2015 at 10h42) and overlay regional boundaries to facilitate locating the Wosret region on the nucleus. The final landing site of Philae, Abydos, is indicated by the yellow arrow. Center: 3D view of the small lobe (always from the same May 2015 data shown on the left side) with  the geomorphological map of Wosret from Lee et al. (2016) superimposed. Right: Sub-units  on top identified by Thomas et al. (2018) in Wosret (Wr a, Wr b, and Wr c) and other regions (Ne, So, An, Gb, and Be refer to Neith, Sobek, Anhur, Geb, and Bes regions, respectively), with the associated comet view from an image acquired on 2 Jan. 2016 at 17h23  (bottom-right). }
\label{wosret}
\end{figure*}

The Rosetta mission of the European Space Agency carried out an in-depth investigation of the short-period comet 67P/Churyumov-Gerasimenko (hereafter 67P) from 1 August 2014 until 30 September 2016, when it ended its operations landing on the comet surface. The Philae lander on board Rosetta was delivered on 12 November 2014 and, after the unexpected rebound on the Agilkia landing site (a second landing at 
30 m away from the final stop that was recently identified; see O'Rourke et al. 2020), it finally landed on Abydos, where most of the foreseen in situ  scientific measurements were performed. \\

The 67P comet has a peculiar bilobate shape  with a surface characterized by extensive layering and showing a variety of terrains including both consolidated and smooth areas, pits, depressions, boulders, and talus (Sierks et al. 2015, Thomas et al. 2015, Vincent et al. 2015, Massironi et al. 2015), of which 26 regions have been defined by structural or morphological boundaries (El Maarry et al. 2015, 2016). \\
Spectroscopy of comet 67P from the Visible, InfraRed, and Thermal Imaging Spectrometer (VIRTIS) reveals a red spectrum featureless until $\sim$2.9 $\mu$m, then showing a broad absorption in the 2.9-3.6 $\mu$m region, indicating a composition dominated by refractory materials rich in organics (Capaccioni et al. 2015, Quirico et al. 2016). Recent recalibration of the VIRTIS instrument permits the identification of some faint absorption bands inside the broad 2.9-3.6 $\mu$m feature, as well as bands that reveal the presence of aliphatic organics (Raponi et al. 2020), and of ammonium salts (Poch et al. 2020).
 
The southern hemisphere regions of the comet became observable by Rosetta only since February 2015, that is about six months before the perihelion passage of the comet. For safety reasons, the spacecraft was kept at relatively high altitude from the nucleus ($>$ 100 km), permitting only low-medium  spatial resolution (i.e. $>$ 2 m/px) investigation of the southern hemisphere regions. These regions were observed at high spatial resolution (sub-meter) only during the Rosetta extended phase in 2016, when the spacecraft approached again to the nucleus as done for the northern hemisphere in 2014 during the Philae landing site selection and characterization (Sierks et al. 2015, Thomas et al. 2015). \\
The data acquired in 2016 permitted a detailed investigation of the physical properties and of the surface evolution due to activity for some southern hemisphere regions. The detailed analysis of Anhur and Khonsu, two regions of the big lobe of the comet,  were reported in Fornasier et al. (2017, 2019a) and Hasselmann et al. (2019), respectively, while that of the Abydos landing site and surroundings was recently published in Hoang et al. (2020). 

In this paper, we aim at investigating Wosret, a southern hemisphere region located in the small lobe of the nucleus (Fig.~\ref{wosret}).  As the other southern hemisphere regions, Wosret experiences intense thermal changes during the orbit because it is illuminated close to the perihelion, when the incoming solar heating on the comet reaches its peak (Marshall et al. 2017). \\
Wosret is a region with peculiar surface morphology: it is dominated by consolidated outcropping material showing lineaments and long fractures, and it looks flattened-out probably because of the high erosion rate associated with the strong thermal effects (El-Maarry et al. 2016). Moreover, it is an active region source of several jets and it includes the second and the last (Abydos) landing site of Philae. Therefore, the investigation of Wosret provides the context of the physical properties measured for the landing site. Finally, covering most of the southern part of the small lobe and being highly eroded, Wosret exposes the inner layers of the small lobe (Penasa et al. 2017). Thus, the comparison of its spectrophotometric properties with those of southern hemisphere regions in the big lobe affected by the same high erosion rate allows for the investigation of surface colors and possible compositional variation between the two lobes.

The article is organized as follows: Section 2 summarizes the OSIRIS observations of Wosret and the methodology applied in our analysis. Section 3 describes the geomorphology of the region and the surface morphological changes. In Section 4, we present the analysis of Wosret spectrophotometric properties and constraints on the water ice content in the localized bright spots identified. Section 5 focuses on the cometary activity events reported for Wosret. Finally, in Section 6 we discuss our findings and we compare them to the published results on comet 67P. This comparison allows us to highlights some differences in the physical properties between the small and big lobe of comet 67P.

\section{Observations and data reduction}

\begin{table*}
\small{
\begin{center}
\caption{Observing conditions of the Wosret region for the NAC images.}
\label{tab1}
\begin{tabular}{|c|c|c|c|c|c|c|} \hline
Day & UT & $\Delta$ & $\alpha$ & res & slope & filters \\
& & (km) & ($^{\circ}$) & (m/px) & \%/(100 nm) & \\ \hline
21/02/2015 & 12h52 & 69.5 & 44.3 & 1.31 & 15.31$\pm$1.62 & F22, F24, F41 \\
25/03/2015 & 02h37 & 99.1 & 77.6 & 1.87 &  16.20$\pm$1.92 & F22, F23, F24, F41  \\
13/04/2015 & 05h59 & 151.8 & 79.2 & 2.86 & 15.87$\pm$1.42 & F22, F24, F41 \\
02/05/2015 & 10h42 & 123.9 & 61.2 & 2.32 & 15.49$\pm$1.50 & all \\
02/05/2015 & 11h42 & 123.9 & 61.1 & 2.32 & 15.44$\pm$1.57 & all \\ 
22/05/2015 & 15h34 & 129.0 & 61.1 & 2.43 & 15.37$\pm$1.60 & all \\
26/07/2015 & 15h10 & 166.9 & 90.1 & 3.14 & 16.78$\pm$1.49 & all \\
01/08/2015 & 07h38 & 215.2 & 90.0 & 4.05 & 16.70$\pm$2.16 & all \\
01/08/2015 & 13h51 & 211.5 & 89.8 & 3.98 & 16.33$\pm$2.11 & all \\
01/08/2015 & 18h13 & 209.6 & 89.8 & 3.95 & 16.48$\pm$2.65 & all \\
22/08/2015 & 22h16 & 329.7 & 88.7 & 6.20 & 15.90$\pm$1.30 & all \\
23/08/2015 & 10h19 & 333.3 & 87.8 & 6.27 & 15.64$\pm$1.62 & all \\
30/08/2015 & 08h09 & 404.7 & 70.3 & 7.62 & 14.94$\pm$1.13 & all \\
30/08/2015 & 12h21 & 404.2 & 70.2 & 7.61 & 15.29$\pm$0.94 & all \\
30/08/2015 & 18h13 & 402.7 & 70.2 & 7.58 & 14.92$\pm$1.15 & all \\
31/08/2015 & 00h34 & 403.1 & 70.2 & 7.59 & 15.34$\pm$0.92 & all \\
11/10/2015 & 18h53 & 524.7 & 61.2 & 9.88 & 14.89$\pm$0.80 & all \\
20/10/2015 & 00h02 & 420.9 & 64.4 & 7.92 & 14.99$\pm$0.86 & all \\
31/10/2015 & 12h46 & 302.0 & 62.8 & 5.68 & 15.32$\pm$0.92 & all \\
31/10/2015 & 16h07 & 297.2 & 62.4 & 5.59 & 15.29$\pm$0.60 & all \\
31/10/2015 & 19h07 & 293.3 & 62.1 & 5.52 & 15.36$\pm$0.76 & all \\
31/10/2015 & 22h49 & 288.4 & 61.7 & 5.43 & 15.25$\pm$0.85 & all \\
19/11/2015 & 20h06 & 125.9 & 78.2 & 2.37 & 16.10$\pm$0.88 & F22, F24, F41 \\
20/11/2015 & 05h16 & 130.0 & 82.3 & 2.45 & 15.78$\pm$1.90 & F22, F24, F41 \\
22/11/2015 & 11h41 & 128.6 & 89.6 & 2.41 & 16.58$\pm$1.24 & F22, F24, F41 \\
22/11/2015 & 19h51 & 126.5 & 89.5 & 2.38 & 17.29$\pm$0.64 & F22, F24, F41 \\
23/11/2015 & 00h13 & 125.7 & 89.5 & 2.37 & 16.65$\pm$1.06 & F22, F24, F41 \\
28/11/2015 & 19h04 & 125.0 & 90.4 & 2.35 & 17.31$\pm$1.70 & all \\
07/12/2015 & 01h13 & 97.9 & 89.7 & 1.84 & 17.07$\pm$1.18 & F22, F24, F41 \\
10/12/2015 & 01h31 & 101.6 & 89.8 & 1.91 & 17.11$\pm$1.48 & F22, F24, F41 \\
12/12/2015 & 23h51 & 99.8 & 89.9 & 1.87 & 16.84$\pm$1.46 & all \\ 
17/12/2015 & 22h56 & 93.4 & 90.4 & 1.75 & 16.59$\pm$2.74 & F22, F24, F41 \\ 
17/12/2015 & 23h56 & 94.0 & 90.4 & 1.76 & 16.90$\pm$1.82 & F22, F24, F41 \\ 
18/12/2015 & 11h09 & 95.1 & 90.5 & 1.78 & 16.76$\pm$2.74 & F22, F24, F41 \\ 
18/12/2015 & 12h09 & 96.0 & 90.5 & 1.80 & 17.01$\pm$2.48 & F22, F24, F41 \\ 
09/01/2016 & 15h05 & 76.5 & 90.5 & 1.44 & 19.14$\pm$2.36 & all \\
23/01/2016 & 17h03 & 74.2 & 62.4 & 1.40 & 16.66$\pm$2.13 & F22, F24, F41 \\
27/01/2016 & 18h20 & 69.0 & 62.8 & 1.30 & 16.49$\pm$1.63 & F22, F23, F24, F27, F28, F41, F71, F61, F16 \\
27/01/2016 & 21h22 & 68.1 & 62.7 & 1.28 & 16.58$\pm$1.01 & F22, F23, F24, F27, F28, F41, F71, F61, F16 \\
10/02/2016 & 19h20 & 46.9 & 65.2 & 0.88 & 17.52$\pm$2.39 & all \\ 
10/02/2016 & 08h06 & 48.9 & 66.3 & 0.91 & 17.68$\pm$1.85 & all \\ 
10/02/2016 & 07h06 & 49.0 & 66.5 & 0.92 & 17.34$\pm$2.17 & all \\ 
09/04/2016 & 17h33 & 32.7 & 40.8 & 0.62 & 16.20$\pm$1.32 & all \\
16/05/2016 & 22h23 & 7.8 & 101.6 & 0.15 & 18.68$\pm$4.79 & F22, F23, F24, F16, F41 \\
28/05/2016 & 12h26 & 4.9 & 101.0 & 0.09 & 19.76$\pm$3.15 & F22, F23, F24, F16, F41 \\
12/06/2016 & 22h28 & 27.5 & 81.4 & 0.52 & 18.13$\pm$2.09 & F22, F24, F41 \\
13/06/2016 & 11h31 & 27.6 & 69.6 & 0.52 & 17.52$\pm$2.10 & F22, F24, F41 \\
13/06/2016 & 23h15 & 27.0 & 61.1 & 0.51 & 17.32$\pm$2.15 & F22, F24, F41 \\ 
14/06/2016 & 10h29 & 26.7 & 54.0 & 0.50 & 17.17$\pm$2.07 & F22, F24, F41 \\
25/06/2016 & 12h16 & 16.1 & 91.4 & 0.30 & 18.28$\pm$1.56 & F22, F23, F24, F16, F41 \\  
25/06/2016 & 13h31 & 15.9 & 93.5 & 0.30 & 18.64$\pm$1.51 & F22, F23, F24, F16, F41 \\ 
02/07/2016 & 15h22 & 14.3 & 92.6 & 0.27 & 18.99$\pm$1.90 & F22, F23, F24, F16, F41 \\ 
03/07/2016 & 12h43 & 7.1 & 102.1 & 0.13 & 18.86$\pm$2.32 & F22, F24, F41 \\
03/07/2016 & 13h03 & 3.0 & 102.3 & 0.06 & 18.80$\pm$3.20 & F22, F24, F41 \\
09/07/2016 & 15h33 & 12.1 & 99.0 & 0.23 & 19.04$\pm$2.69 & F22, F23, F24, F16, F41, F27, F71 \\ 
09/07/2016 & 16h03 & 12.0 & 98.5 & 0.22 & 19.11$\pm$2.59 & F22, F23, F24, F16, F41, F27, F71 \\ 
09/07/2016 & 16h33 & 11.9 & 97.8 & 0.22 & 19.00$\pm$2.94 & F22, F23, F24, F16, F41, F27, F71 \\ 
16/07/2016 & 16h39 & 9.3 & 105.7 & 0.17 & 18.43$\pm$3.11 & F22, F23, F24, F16, F41, F27, F71 \\ 
24/08/2016 & 18h17 & 3.9 & 91.5 & 0.07 & 18.52$\pm$3.06 & F22, F24, F41 \\
\hline
\end{tabular}
\end{center}
}
\tablefoot{$\alpha$ is the phase angle, $\Delta$ is the distance between comet and spacecraft, and ``res" the spatial resolution. The time refers to the start time of the first image of an observing  sequence. Filters: F22 (649.2 nm, reference), F23 (535.7 nm), F24 (480.7 nm), F16 (360.0 nm), F27 (701.2 nm), F28 (743.7 nm), F41 (882.1 nm), F51 (805.3 nm), F61 (931.9 nm), F71 (989.3 nm), and F15 (269.3 nm). The spectral slope was evaluated in the 535-882 nm range after normalization at 535 nm using the F23 filter image when present, and otherwise an artificial 535 nm image was created from the interpolation of the observations acquired at 480 nm and 649 nm.}
\end{table*}

\begin{figure*}
\centering
\includegraphics[width=1.0\textwidth]{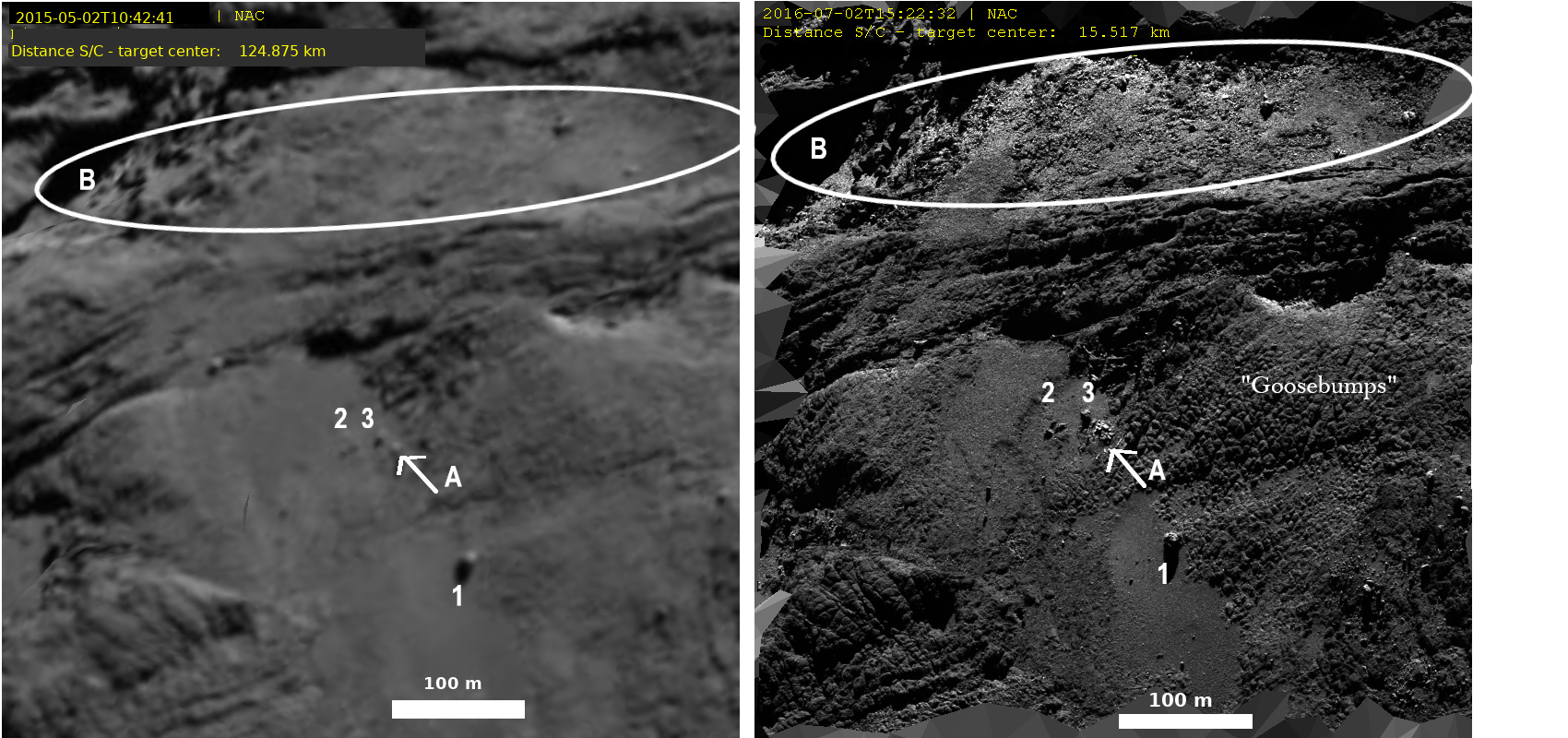}
\caption{Comparison between images acquired on May 2015 (on the left, spatial resolution ({\it res}) of 2.3 m/px) and on July 2016 (on the right, res = 0.3 m/px) covering part the Wosret region (area {\it a} following Thomas et al. (2018) sub-regions definition). A new cluster of outcrops (named {\it A} and indicated by the white arrow) is clearly visible on the July 2016 images. Some boulders are indicated for reference (and numbered 1, 2, and 3). The ellipse named B includes a region where the dust coating may also have been thinned during the perihelion passage, as observed close to the cluster of outcrops named {\it A} and to the goosebump features.}
\label{morphoev1}
\end{figure*}

For the purposes of this study, we used data from the Optical, Spectroscopic, and Infrared Remote Imaging System (OSIRIS) on board Rosetta. This imaging system is composed of a Narrow Angle Camera (NAC), which was mostly devoted to the investigation of the 67P nucleus, and of a Wide Angle Camera (WAC) which investigated the wide-field coma (Keller et al. 2007).\\
Wosret was observed only since February 2015 and until the end of the Rosetta mission (Table~\ref{tab1}), except for small areas close to the Maftet, Hatmehit and Bastet boundaries that were visible also before. Because of the high activity phase of the comet and of the associated higher altitude orbits of Rosetta during most of the 2015 observations, Wosret was observed at sub-meter spatial resolution only in 2016, during the extended phase of the Rosetta mission. \\
The images analyzed here come from the NAC camera, which was equipped with 11 filters covering the 250-1000 nm range. We searched in the OSIRIS archive, which includes more that 75000 images of the comet, for all the NAC spectrophotometric sequences covering the Wosret region and having at least three filters. \\
We used radiance factor (also known as $I/F$) data produced by the OSIRIS standard pipeline after the bias, flat-field, geometric distortion,  and absolute flux calibration steps corrections described in Tubiana et al. (2015):
\begin{equation}
Radiance Factor (i,e,\alpha,\lambda)  =  \frac{\pi I(i,e,\alpha,\lambda)}{F_{\lambda}},
\end{equation}
where I is the observed scattered radiance, $F_{\lambda}$ the incoming solar irradiance at the heliocentric distance of the comet and at a given wavelength ($\lambda$), and  $i$, $e,$ and $\alpha$ are the incidence, emission, and phase angles, respectively.

To perform the spectrophotometric analysis, the images of a given sequence were first coregister using the F22 NAC filter (centered at 649.2nm) as a reference and applying  devoted python scripts, as done previously in similar studies of 67P comet (Hasselmann et al. 2019, Fornasier et al. 2019a). 
Then the Lommel-Seeliger disk function ($D(i,e)$) was applied to correct the illumination conditions:
\begin{equation}
D(i,e) = \frac{2\mu_{i}}{\mu_{e}+\mu_{i}}
,\end{equation}
where $\mu_{i}$ and $\mu_{e}$ are the cosine of the solar 
incidence  and emission angles, respectively.  
To retrieve the illumination conditions we used the 3D stereophotoclinometric (SPC, SHAP8 version) shape model of the 67P nucleus (Jorda et al. 2016). The RGB images were generated using the filters centered at 882 nm (R), 649 nm (G), and 480 nm (B), and the STIFF code (Bertin 2012).\\
To compute the spectral slope, we first normalized the data at 535 nm (filter F23), then we  computed it in the 882-535 nm wavelength range to be consistent with the methodology applied for 67P spectrophotometry in previous publications. When the F23 filter was not in the observing sequence, we produce an artificial image at 535 nm through linear interpolation of the images acquired at 480 nm and 649 nm. \\
The average spectral slope reported in Table~\ref{tab1} was evaluated fitting the slope distribution with a Gaussian function, where the average spectral slope is derived from the peak value of the Gaussian and the associated error from the standard deviation of the fit. For the medium-low resolution data covering also other regions, the slope maps were first cut selecting the area covering Wosret before applying the Gaussian fit.

Details on the observing conditions and the Wosret spectral slope values are reported in Table~\ref{tab1}. \\

\begin{figure*}
\centering
\includegraphics[width=0.99\textwidth]{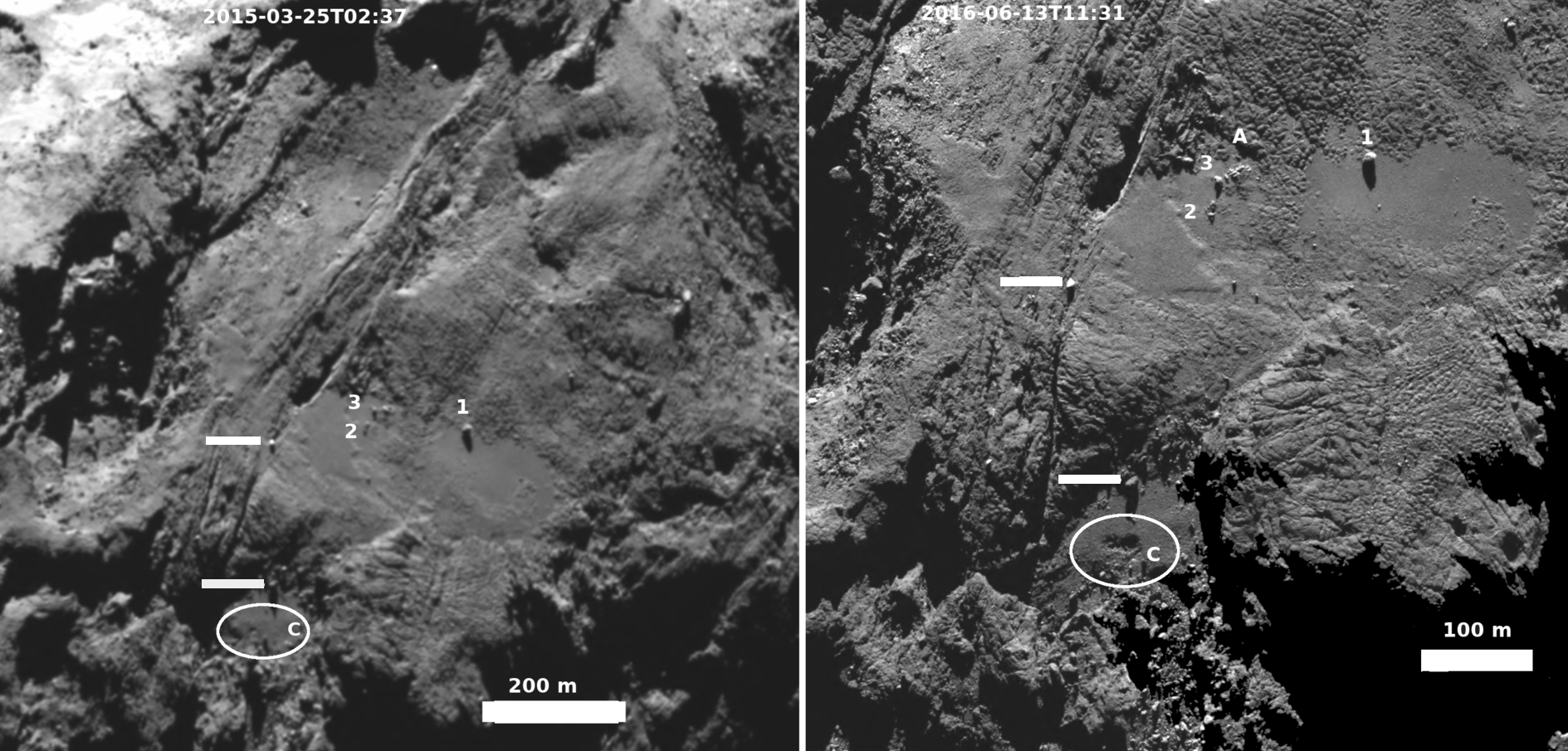}
\caption{Comparison between images acquired on March 2015 (left side, res = 1.9 m/px) and June 2016 (right side, res = 0.5 m/px) showing the formation of a new cavity inside the ellipse named {\it C}. Some boulders are indicated for reference by white bars and numbers, following the same numbering scheme of Fig.~\ref{morphoev1}.} 
\label{cavity}
\end{figure*}

To estimate the height $h$ of a given surface feature, we considered its projected shadow and the associated geometric conditions, determined by  ancillary images produced using an image simulator (Hasselmann et
al. 2019) supplied with pre-calculated NAIF SPICE Kernels\footnote{\url{https://www.cosmos.esa.int/web/spice/spice-for-rosetta}}
and the SPC shape model (Jorda et al. 2016). Any misalignment
between the ancillary images and the observed one was corrected through
translation and rotation transformations calculated from uniquely
identified tie landmarks in both images. We thus used the following equation to determine the height ({\it h}) of a given surface feature:
\begin{equation}
h= L_{sha}\cdot\tan(\pi/2-i), 
\label{height}
\end{equation}
where $L_{sha}$ is the length of the shadow of the feature, and i the incident angle. 
The incidence angle here considered is the angle between the vector centered in the tip
of the shadow, connecting the top of the structure, and the normal
vector to the surface of the feature (see Fig. 2 from Cambianica et al. 2020). \\
This method has been tested and applied in many analyses of meter-sized features on the OSIRIS
images of the nucleus to estimate local surface changes (El-Maarry et al. 2017, Hasselmann et al. 2019, Cambianica et al. 2020).

 \section{Geomorphology and surface evolution}


\begin{figure}
\begin{centering}
\includegraphics[width=0.47\textwidth,angle=0]{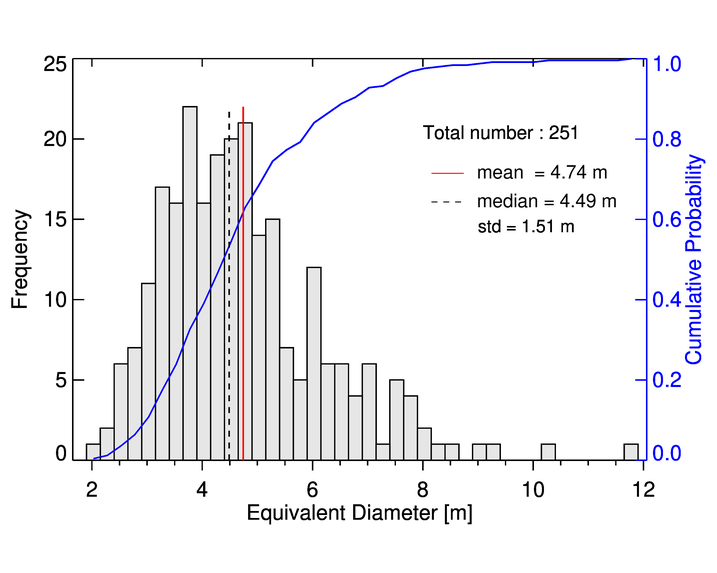}
\par
\end{centering}
\caption{Goosebumps equivalent diameter from the 2 July 2016 image (see Fig.~\ref{morphoev1})}
\label{goosebumps}
\end{figure}

Wosret is surrounded by Maftet and Bastet in the West and East sides, respectively, by the Hatmehit depression at North, and by the Neith region at South (Fig.~\ref{wosret}). 
Wosret is dominated by consolidated outcropping material showing ubiquitous fractures up to $\sim$ 300 m in length, lineaments (El-Maarry et al. 2016), and  some pits and niches observed to be periodically active during the perihelion passage (Vincent et al. 2016, Fornasier et al. 2019b).
The region has a surface of about 2.36 km$^2$ covering most of the small lobe’s southern hemisphere, and it has been divided by Thomas et al. (2018) into three areas (Wr a, Wr b, and Wr c), showing topographic and textural difference (Fig.\ref{wosret}). Area {\it Wr a} is relatively flat and smooth, {\it Wr b} is dominated by fractured terrains, while area {\it Wr c} shows a higher roughness texture and it includes almost circular depressions and ridges (Thomas et al. 2018). As shown in Fig.~\ref{wosret},   Wosret lacks of wide-spread dust coatings, except for few pits filled by smooth materials (El-Maarry et al. 2016). 

Wosret has a peculiar flattened-out aspect (El-Maarry et al. 2016), probably because of the high erosion rate associated with the strong thermal effects. In fact, Keller et al. (2015, 2017) estimated an erosion rate up to  four  times higher in the southern hemisphere than in the northern one. Concerning Wosret,  Lai et al. (2016) estimated a dust loss up to 1.8 m in depth per orbit. Wosret is also the region which has the highest water production rate during perihelion, according to the Microwave Instrument for the Rosetta Orbiter (MIRO) measurements (Marschall et al. 2017).  

Figures~\ref{morphoev1} and ~\ref{cavity} show part of the Wosret region (sub-unit {\it Wr a} following Thomas et al. (2018) classification) observed in 2015, before the perihelion passage and in 2016. The images taken on March and May 2015 have not sufficient spatial resolution (2.3 m/px) nor good illumination conditions ($i\sim30^{\circ}$) to accurately measure the typical depth of surface features from their shadows. Therefore, it is not possible to precisely quantify the possible evolution of the dust cover in this area through the perihelion passage. 

\begin{figure*}[t]
\centering
\includegraphics[width=0.97\textwidth]{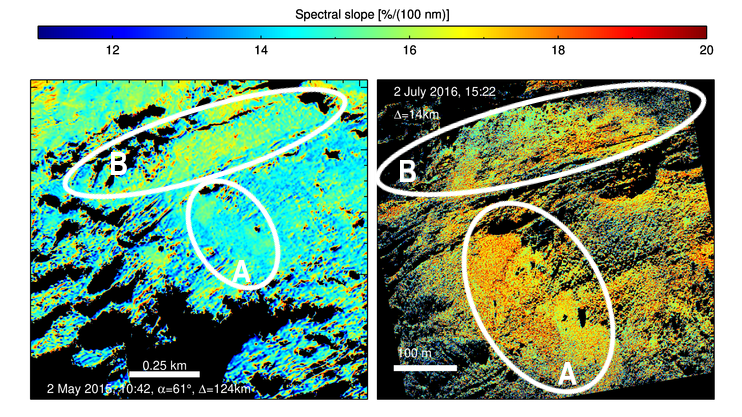}
\caption{Spectral slope evolution with heliocentric distance comparing data acquired on 2 May 2015 (r$_h$ = 1.73 AU inbound) and 2 July 2016 (r$_h$ = 3.32 AU outbound) images. The July 2016 data were phase reddening corrected from 93$^o$ to 62$^o$ using the phase reddening coefficient of 0.055 $\times 10^{-4} nm^{-1} deg^{-1}$ determined in section~\ref{phasereddeningcoeff} for Wosret in the post-perihelion 2016 observations. Comparing to the original data, we applied a zoom factor of 2 to the May 2015 data, while the July 2016 image was binned by a factor of 2.} 
\label{slopeevolution}
\end{figure*}
Conversely, the NAC images acquired on June and July 2016, have a high spatial resolution (0.27-0.52 m/px) which highlights some interesting features such as highly fractured terrains showing polygonal block patterns (F2 structure in El-Maarry et al. 2016). The texture reminds the surface features referred to as "goosebumps"{} (see Fig.~\ref{morphoev1}) or clods, first noticed on the 67P comet inside the Seth pits (Sierks et al. 2015), but later observed in other regions. The Wosret blocks appear consolidated, highly irregular in shape and height. We estimate the size of 251 blocks from the 2 July 2016 image, finding an equivalent diameter ranging from 2 to $\sim$12 m, with an average value of 4.74$\pm$1.51 m (Fig.~\ref{goosebumps}). Wosret clods are larger than the average diameter values ($<D>$) reported for similar structures observed in the big lobe of the comet, such as Seth ($<D>$ = 2.2 m, Sierks et al. 2015), Imhotep and Anubis ($<D>$ = 3.2 m and 2.5 m, respectively, Davidsson et al. 2016), or Atum, where the average diameter is 2.2 m but, in its southern part, some clods are larger, with size ranging from 3.2--4.6 m (Davidsson et al. 2016).

\begin{figure*}
\centering
\includegraphics[width=0.97\textwidth]{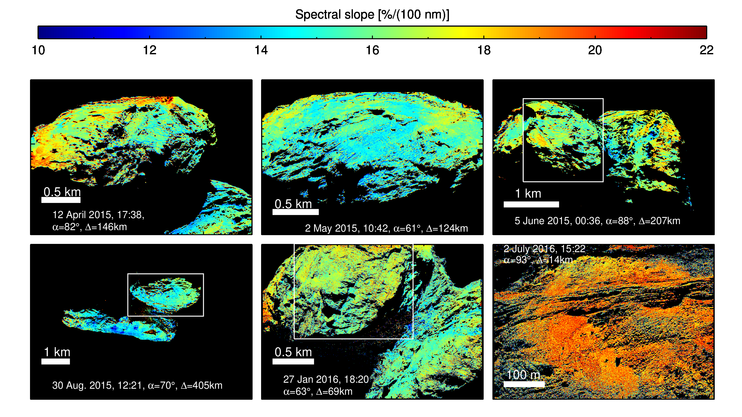}
\caption{Spectral slope of the Wosret region, evaluated in the 535-882 nm range, from different images in the April 2015-July 2016 time frame. The Wosret region is indicated by white rectangles in the lower resolution images. The high resolution image acquired on July 2016 was binned by a factor of 2.} 
\label{slopes}
\end{figure*}
In comparing the pre- and post-perihelion images, we also identify a couple of clear morphological changes: a cluster of bright outcrops at $lat={-15.15}^{\circ}$, and $lon={343.2}^{\circ}$ (feature A in Fig.~\ref{morphoev1}); a new cavity at $lat =-19.7^{\circ}$, and $lon=335.9^{\circ}$ (inside the ellipse called C in Fig.~\ref{cavity})

\subsection{Outcrops cluster}

The cluster of outcrops is clearly visible in the July 2016 image (Fig.~\ref{morphoev1}), and it includes 15 outcrops having 2 to 5 meters in 
size. In the 2 July 2016 image acquired at 15h22, the outcrops look brighter than the surroundings. This is not, however, an intrinsic effect related to a different composition, but simply an illumination effect. In fact, OSIRIS also acquired an image of the same area one hour before the one presented in Fig.~\ref{morphoev1}, and there the brightness of the cluster is similar to that of the average terrain. In the 2015 images, only one block of the cluster is visible, while most elements of the cluster are buried beneath the dust. As said before, the resolution of the Wosret pre-perihelion images is too low and our attempts to estimate the features height were unsuccessful, thus, we cannot precisely quantify the dust coating variation. However, we were able to measure the height of the outcrops  with shadows clearly ending on the background terrain in the 2016 images, finding an average height of $1.6\pm0.3$ meters. We can thus infer that the thickness of the dust coating in this area diminished by $\sim$ 1 meter from pre-perihelion to post-perihelion to reveal the 15 outcrops of the cluster. We visually inspected and compared the pre-perihelion images and the post-perihelion ones binned by a factor of 8 to have a comparable spatial resolution between the two datasets. The pre-perihelion images look smoother in several areas, including the one around the outcrops, showing the goosebumps features, and the area at the northern side of Wosret, indicated by the ellipse named {\it B} in Fig.~\ref{morphoev1}. We thus may tentatively deduce that the dust coating was thinned in the aforementioned areas too.\\
Dust movements in this region are also supported by the observed colors variation reported in Fig.~\ref{slopeevolution}, where the spectral slope of the same  pre- and post-perihelion images presented in Fig.~\ref{morphoev1} are shown. As these images were acquired at different heliocentric distances and phase angles, and considering that comet 67P has a strong spectral phase reddening effect, the July 2016 image was phase reddening corrected from phase angle 93$^o$ to 62$^o$ using the Wosret phase reddening coefficient here determined from 2016 observations (see section~\ref{phasereddeningcoeff}), in order to match the viewing geometry of the May 2015 image. 
As indicated in Fig.~\ref{slopeevolution}, the region included inside the {\it A} ellipse was relatively spectrally bluer (i.e., it has a lower spectral slope)  than that included in the {\it B} ellipse in pre-perihelion data. Conversely, in July 2016 data the region included inside the {\it A} ellipse looks redder or as red as the one inside the {\it B} ellipse, indicating local surface evolution related to the composition or to different grain size or roughness of the material.

\subsection{New cavity}

The cavity that formed between March 2015 and June 2016 has
apparently expanded from another small depression visible on 25 March 2015 image. 
It is $\sim$ 30 m long and 11-15 m large, having an area of 492$\pm$32 m$^2$. 
We traced seven shadow lengths throughout
the cavity, from which we derive a depth of $6.5\pm0.8$ meters.
Assuming for the loss material the density of the bulk nucleus ($537.8\pm2 ~kg/m^{3}$ (Patzold et al. 2016, Jorda et al. 2016, Preusker et al. 2017), we estimate that (1.2$\pm$0.2$)\times$10$^{6}$ kg of cometary material was lost in this cavity. For comparison, this mass is similar to that lost in a couple of new scarps (or cliffs) observed in the Anhur region (see Figs. 2 and 10 in Fornasier et al. 2019a). However, Anhur, as well as the Khonsu region or the Aswan site, also shows some drastic surface changes with an estimated mass loss at least ten times larger than that of the cavity here reported (Fornasier et al. 2019a, Hasselmann et al. 2019).


%

\section{Spectrophotometry and bright spots}

\begin{table*}
         \begin{center} 
         \caption{Phase reddening coefficients evaluated for Wosret during different observations and those reported in the literature for comet 67P and for other dark small bodies. }
         \label{tab_reddening}
        \begin{tabular}{|l|l|c|c|l|} \hline
Body and conditions & Wavelength & $\gamma$ & Y$_0$ & References \\
                    & range (nm) & ($10^{-4}$ nm$^{-1} deg^{-1}$) & ($10^{-4}$ nm$^{-1}$)  & \\ \hline
{\bf Comet 67P}           &            &                               &               & \\ 
Wosret post-perihelion (2016) & 525-882   &  0.0546$\pm$0.0042 & 13.6$\pm$0.4 & this work \\
Wosret around perih. (2015) &   525-882   & 0.0396$\pm$0.0067  & 12.7$\pm$0.5    & this work \\
North. hemisp. pre-perih (2014)   & 525-882   & 0.1040$\pm$0.0030  & 11.3$\pm$0.2 & Fornasier et al. 2015 \\
North\&South hemisph. at perih. (2015) & 525-882   & 0.0410$\pm$0.0120  & 12.8$\pm$1.0 &  Fornasier et al. 2016 \\
Small region in Imhotep, Ash, Apis (Feb. 2015 Fly-by)  & 535-743 & 0.0652$\pm$0.0001  & 17.9$\pm$0.1 & Feller et al. 2016 \\
Abydos and surroundings (post-perihelion, 2016) & 535-882 & 0.0486$\pm$0.0075  & & Hoang et al. 2020 \\
Northern hemisph. pre-perih. (July 2014-Feb 2015)              & 1000-2000  &0.018 & 2.3 & Ciarniello et al. 2015 \\
Northern hemisph. pre-perih. (2014)              & 1100-2000  &0.015$\pm$0.001 &  2.0  &Longobardo et al. 2017 \\ \hline
D-type asteroids                                  &450-2400    & 0.05$\pm$0.03  &       & Lantz et al., 2017 \\
(1) Ceres                                            & 550-800    &  0.046         & -0.2 & Ciarniello et al. 2017\\
(101955) Bennu                                  & 550-860 & 0.014$\pm$0.001 & -1.3$\pm$0.1 & Fornasier et al. 2020 \\ 
(162173)  Ryugu                                 & 550-860    & 0.020$\pm$0.007 &  & Tatsumi et al. 2020 \\  \hline
        \end{tabular}
\end{center}
\tablefoot{The quantity $\gamma$ is the phase reddening coefficient; Y$_0$ is the estimated spectral slope at zero phase angle from the linear fit of the data.}
 \end{table*}

Comet 67P shows compositional heterogeneities which have been observed both with the VIRTIS spectrometer and the OSIRIS cameras (Fornasier et al. 2016,  Barucci et al. 2016, Filacchione et al. 2016b, Deshapryia et al., 2016; Oklay et al. 2017). From OSIRIS pre-perihelion data Fornasier et al. (2015) reported spectral slope variations from $\sim$ 11 to 21 \%/(100 nm) in the 535-882 nm range and at phase $\sim$ 50$^o$, distinguishing three kind of terrains across the nucleus, from the spectrally bluer and water ice enriched terrains observed in Hapi, to the redder ones, associated mostly with dusty regions. A similar color variation has been reported later for the southern hemisphere (Fornasier et al. 2016). Looking at Table~\ref{tab1} and  Fig.~\ref{slopes}, Wosret is shown to have an average slope of  15.5 \%/(100 nm) at $\alpha= 61^o$ (pre-perihelion), belonging thus to the intermediate color group \# 2, according to Fornasier et al. (2015) spectral classification of 67P nucleus terrains.

\subsection{Spectral phase reddening}
\label{phasereddeningcoeff}
The variation of the spectral slope with phase angle, knows as the spectral phase reddening effect, is the results of small-scale surface roughness and multiple scattering in the surface medium at extreme geometries (high phase angles) and it is commonly observed on several Solar System bodies. This has been reported for asteroids (see Fornasier et al. 2020 and references therein), comets (Fornasier et al. 2015, Ciarniello et al. 2015, Longobardo et al. 2017), as well as planets and their satellites (Gehrels et al. 1964, Warell \& Bergfors, 2008, Nelson et al. 1987, Cuzzi et al. 2002, Filacchione et al. 2012).  In case of a low-albedo surface, such as that of cometary nuclei, the spectral phase reddening effect has been interpreted as due to the presence of fines, namely, particles of $\sim$ micron size, or to the irregular surface structure of larger grains, having micron-scale surface roughness (Li et al. 2019, Schr\"oder et al. 2014, Pilorget et al. 2016). 

Comet 67P has a strong spectral phase reddening effect (Fornasier et al. 2015, 2017, Ciarniello et al. 2015, Longobardo et al. 2017), which is, moreover not fixed over time but it shows seasonal variations depending on the cometary solar distance and activity (Fornasier et al. 2016).  In fact,  the spectral slope value was found to be smaller close to the perihelion compared to the one measured pre- and post-perihelion at similar phase angles. This less red color was interpreted as a consequence of the cometary activity peak during perihelion, with the partial removal of the dust mantle and the consequent exposure of the volatile enriched material underneath. The 67P phase reddening coefficient is monotonic and wavelength dependent (Table~\ref{tab_reddening}), with values of 0.04-0.1 $\times 10^{-4}$ nm$^{-1} deg^{-1}$ in the 535-882 nm range (Fornasier et al. 2015, 2016), and lower values, 0.015--0.018  $\times 10^{-4}$ nm$^{-1} deg^{-1}$ in the 1 to 2 $\mu$m range (Ciarniello et al. 2015, Longobardo et al. 2017).

\begin{figure}
\includegraphics[width=0.5\textwidth,angle=0]{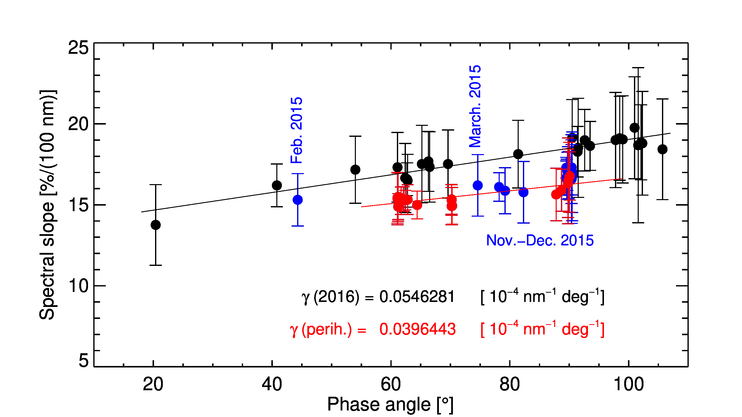}
\caption{Phase reddening in spectral slopes evaluated in the 535-882 nm range. The black circles represent data acquired in 2016, red circles those acquired in May-Oct. 2015, that is three months before and after the perihelion passage, while the remaining 2015 data are represented with blue circles. The black and red lines are the linear fit of the 2016 and near perihelion data, respectively.}
\label{gamma}
\end{figure}

\begin{figure*}
\centering
\includegraphics[width=0.85\textwidth]{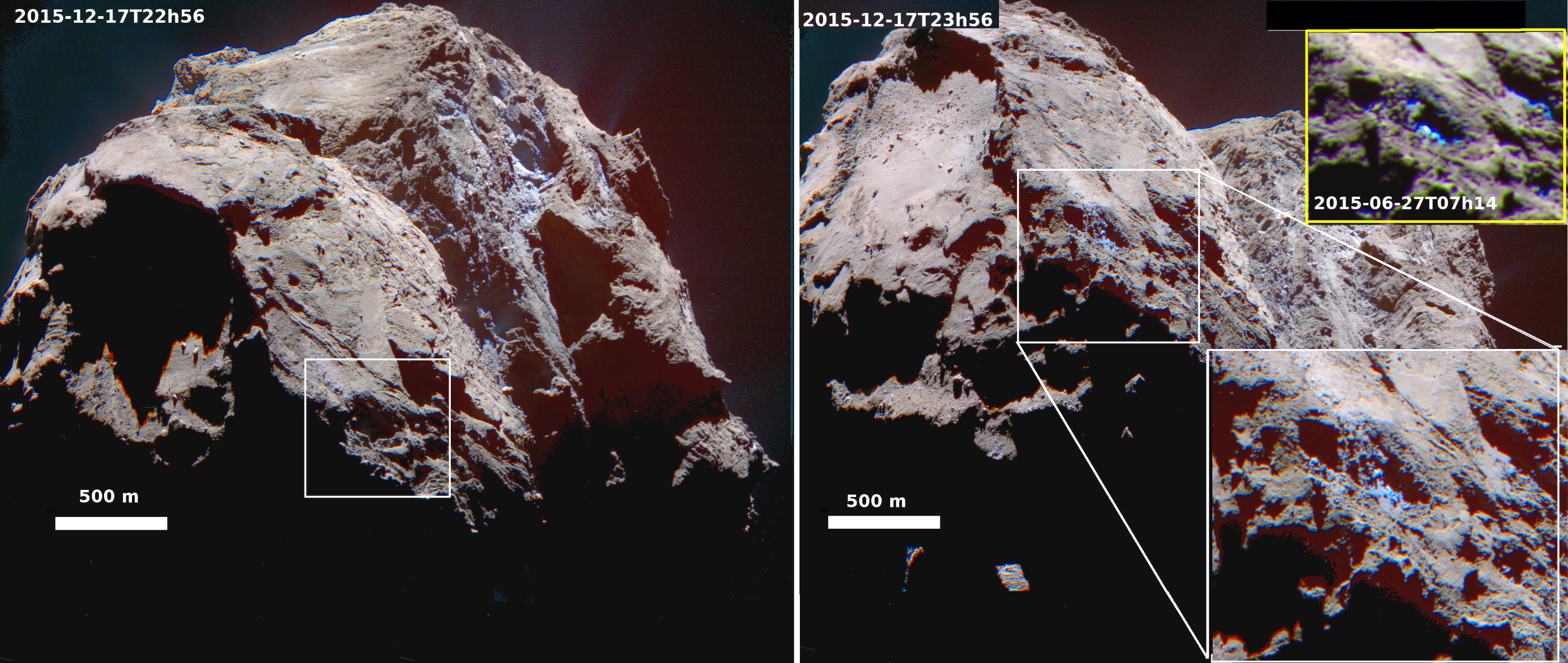}
\caption{RGB images acquired on 17 December 2015 taken one hour apart. The right image acquired at UT23h56  shows a relatively blue region (indicated by the white rectangle) at dawn. The same area was observed in a sequence acquired one hour later (on 18 Dec. 2015, UT 00:56) and it shows a color indistinguishable from that of the average Wosret region. The bluing at dawn is periodic and observed in other sequences in December 2015 but also in June 2015, as shown in the top insert of the image on the right side.} 
\label{blue}
\end{figure*}

As the whole nucleus, also the Wosret region shows changes in the spectral slope with the phase angle (Figs~\ref{slopes} and ~\ref{gamma}, and Table~\ref{tab1}). 
From the reported spectral slopes at different phase angles, we computed the phase reddening coefficient ($\gamma$) and the spectral slope at
$0^{\circ}$ phase angle ($Y0$) applying a linear fit to the data. Because of the different heliocentric distances and cometary activity levels, we evaluated the  spectral reddening coefficient for data acquired in 2016 and for those acquired close to the perihelion passage during the May-October 2015 timeframe (Fig.~\ref{gamma}). Unfortunately, the Wosret region was not observable during 2014, except for tiny fractions, and no data are available at a low phase angle. \\
For the Wosret region, we observe a much shallower phase reddening coefficient compared to the results reported in Fornasier et al. (2015). For the post-perihelion observations of 2016, we find for Wosret a $\gamma_{2016}$ = (0.05462$\pm$0.0042) $\times 10^{-4}$ nm$^{-1} deg^{-1}$, which is about a factor of 2 lower than that 
determined for the northern hemisphere regions in inbound orbits (Fornasier et al. 2015). For the 2015 observations acquired close to the perihelion passage the phase angle coverage is very limited, and the associated phase reddening coefficient ($\gamma_{perih}$) is  (0.0396$\pm$0.0067) $\times 10^{-4}$ nm$^{-1} deg^{-1}$, a value very close to the one found globally for the comet during the perihelion passage (Table~\ref{tab_reddening}). 

Similar trend with progressive lower phase reddening and lower spectral slope values approaching  perihelion were also observed elsewhere on comet 67P:  in the northern hemisphere area (at 5-14$^{\circ}$ latitude) across the Imhotep-Ash regions ($\gamma = 0.06 \times 10^{-4} nm^{-1} deg^{-1}$) during the February 2015 Rosetta flyby (Feller et al. 2016); in the southern hemisphere Anhur region (Fornasier et al. 2017, 2019b); and in the Abydos final landing site of Philae (Hoang et al. 2020), which is located in Wosret but close to the boundaries with the Bastet and Hatmehit regions. For Abydos, Hoang et al. (2020) found a phase reddening coefficient value (Table~\ref{tab_reddening}) very close to the one here determined for the whole Wosret region.  

The phase reddening coefficient determined for the Wosret region is comparable to the one found for other dark and presumed organic rich bodied such as D-type asteroids (0.05$\pm$0.03 $\times 10^{-4}$ nm$^{-1} deg^{-1}$ in the 0.45-2.45 $\mu$m range, Lantz et al. 2017), and the dwarf planet Ceres ($\gamma = 0.046 \times 10^{-4} nm^{-1} deg^{-1}$, Ciarniello et al. 2017, 2020). It is however two to four times higher than the value reported for low albedo and carbonaceous rich bodies such as the near-Earth asteroids: (101955) Bennu and (162173) Ryugu (Table~\ref{tab_reddening}), that were recently visited by the OSIRIS-REx and Hayabusa 2 missions, respectively. \\
Considering that the surfaces of Bennu and Ryugu are dominated by boulders and display a lack of extended fine dust layers (Lauretta et al. 2019, Jaumann et al. 2019), as well as the fact that the 67P phase reddening decreases approaching perihelion, when the activity lifts-up part of the dust mantle, and that Wosret has a lower phase reddening coefficient than the global 67P nucleus, 
we may deduce that the surface structure of Wosret has an intermediate roughness level between that of dust poor surfaces, such as those of Bennu and Ryugu, and the dust-rich ones such as most of the 67P northern hemisphere regions.
The lack of widespread dust coating observed in the southern hemisphere regions of the comet (El-Maarry et al. 2016; Thomas et al., 2018), including Wosret, is probably responsible of its reduced surface micro-roughness relative to most of the northern hemisphere regions of the comet. \\
This is corroborated by the results of Keller et al. (2015, 2017), who found that the dust particles ejected from the southern hemisphere of comet 67P during the peak of activity partially fall back and are then deposited in the northern hemisphere. Thus, the southern hemisphere regions become progressively depleted of dust.

\begin{figure}
\includegraphics[width=0.4\textwidth,angle=0]{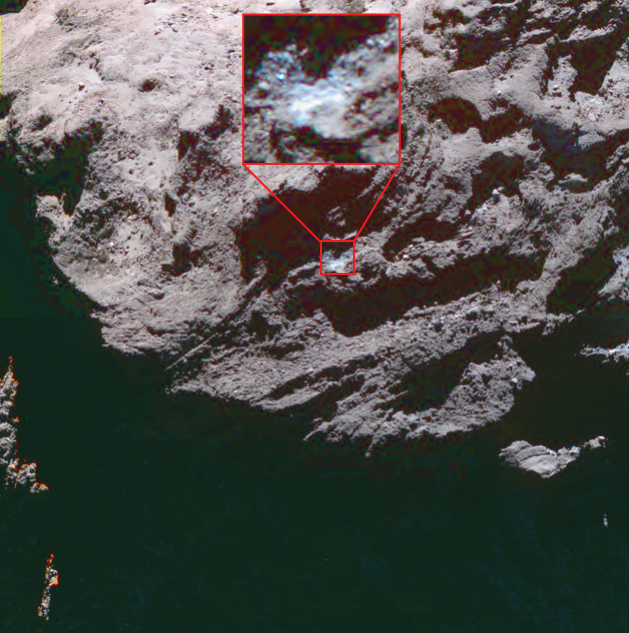}
\caption{RGB image taken on 12 June 2016 at 12h48 showing a blue patch at latitude -21$^o$ and longitude 318$^o$ in an area observed to be active by Vincent et al. (2016, jet numbered 11 in their paper).}
\label{bluepatch}
\end{figure}

\subsection{Water ice exposures}

\begin{figure*}
\centering
\includegraphics[width=0.9\textwidth]{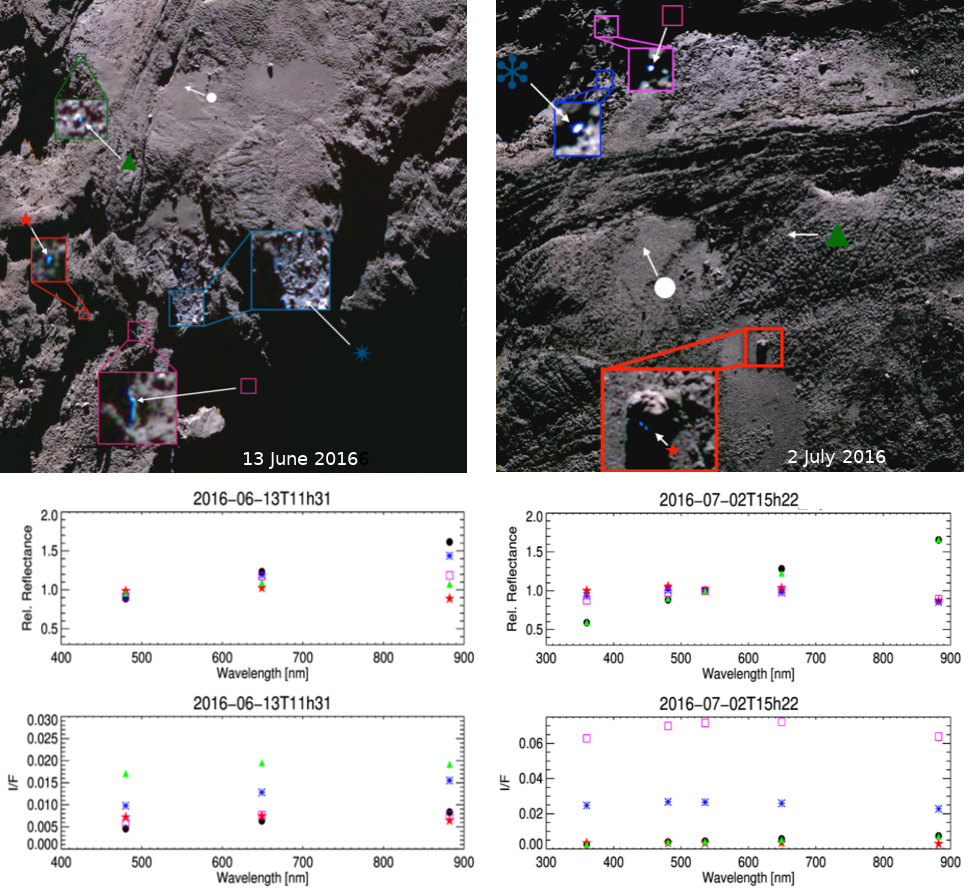}
\caption{Wosret spectrophotometry. Top: Examples of blue patches and bright spots from 13 June 2016 (left) and 2 July 2016 images (right). Inserts show a zoom of some ROIs.  Middle and bottom panels: Corresponding relative reflectance and the I over F value at the given phase angle (70$^o$ and 93$^o$ for the June and July images, respectively). The dark terrain used in the mixing model is represented by a circle in white for clarity in the RGB images on the top and in black in the plots. } 
\label{spotROI}
\end{figure*}

In our spectral analysis of the Wosret region, we notice some local compositional heterogeneity : i) a few areas that look bluer in colors than their surroundings; ii) some localized bright spots showing evidence of exposed volatiles.

Figure~\ref{blue} shows RGB images acquired on 17 December 2015 separated by 1 hour. On the right side of Fig.~\ref{blue} an area (indicated by the white rectangle) emerging from the shadows shows a bluer color than its surroundings. Further images acquired 1-2 hours later, when the region is fully illuminated by the Sun, reveal colors and slope as red as the surroundings.  This diurnal color variation -- namely, a relatively bluer slope at dawn or for an area just emerging from shadows followed later by a color as red as the surrounding terrains when the area is illuminated by the Sun -- is periodic. The diurnal color variation is observed in different data acquired in December 2015, but also before and after the perihelion passage, for example on June 2015, as shown in the insert in Fig.~\ref{blue}, and on June 2016 (Fig.~\ref{bluepatch}). Another example of bluer area  is shown in Fig.~\ref{spotROI} from June 2016 data (the region inside the blue rectangle in the left side image).

This diurnal color variation has been noted at dawn on different areas on both lobes of the comet (Fornasier et al. 2016), and associated  with a surface enrichment of water ice-frost condensed during the previous night, which results in a relatively blue color at dawn. During the comet morning the increasing temperature produces the sublimation of the volatiles and, consequently, the surface gets redder in color because of depleted in water ice-frost. Previous studies made with OSIRIS and VIRTIS have in fact demonstrate that the relatively blue colors of a given area of comet 67P are associated with a local enrichment in volatiles (de Sanctis et al. 2015, Barucci et al. 2016, Fornasier et al. 2015, 2016, Filacchione et al. 2016a, 2016b).

Beside local bluing, we also observe the presence of tiny bright spots (BS), having a surface of 1-2 m$^2$ usually located close to shadows. For these regions of interest (ROIs), we computed the relative spectrophotometry and radiance by integrating them over a box of 3$\times$3 pixels, as reported in Fig.~\ref{spotROI}. These features, besides being brighter than the comet dark terrain, have a distinct spectrophotometric behavior characterized by a low to neutral slope, indicating localized exposures of water ice. For the bright spots we attempt to estimate their water-ice content following the method described in Fornasier et al. (2019), namely, using a simple areal mixing model with two components: the cometary dark terrain (represented by the circle in Fig.~\ref{spotROI}) and water ice
\begin{equation}
R =  p \times R_{ice} + (1-p) \times R_{DT}
 ,\end{equation}
where $R$ is the reflectance of the bright patches, R$_{ice}$ and R$_{DT}$ are the reflectance of the water ice and of the cometary dark terrain, respectively,  and p is the relative surface fraction of water ice. The water-ice spectrum was derived from the synthetic reflectance from Hapke modeling starting from optical constants published in Warren and Brandt (2008) and adopting a grain size of 30 $\mu$m-100 $\mu$m, as is typical  for ice grains on cometary nuclei (Sunshine et al. 2006, Capaccioni et al. 2015, Filacchione et al. 2016a), but also testing larger grain size of 1000 $\mu$m.
Before applying the linear mixing model, the images were photometrically corrected using the Hapke model parameters determined by Fornasier et al. (2015, see their Table 4) to obtain an estimation of the normal albedo of the different ROIs.
In doing so, we assume that the phase function, determined by Fornasier et al. (2015) from the comet surface at 649 nm also applies to the other wavelengths and to the volatile rich spots. This introduces some uncertainties in the water-ice abundance estimation, but because of the short visibility of the bright spots and of the limited phase angle coverage,  it is impossible to determine Hapke photometric parameters for the Wosret region or for the BS on comet 67P. \\
We report in Table~\ref{tab_model} the coordinated of the ROIs investigated, the associated spectral slope, and the estimated water ice content that was determined for BS, having  an incidence and emission angle lower than 75$^o$, in line with the illumination conditions for which the photometric corrections are reliable (Hasselmann et al. 2017). \\
\begin{table*}
         \begin{center} 
         \caption{Water-ice content estimates for the bright features observed in the 13 June and 2 July 2016 images. }
         \label{tab_model}
        \begin{tabular}{|llllll|} \hline
time          &  ROI & lon & lat & Spectral slope & Ice content (30,100,1000 $\mu$m)  \\ 
              &      & [$^o$] & [$^o$] & [\%/100nm]     &   (\%)                     \\ \hline

2016-06-13T11h31 & circle          & 340.28   &  -14.94  &    17.72 & 0, DT     \\
2016-06-13T11h31 & red star        & 327.72  &   -21.09 &    -4.73 &  in shadows        \\
2016-06-13T11h31 & blue star       & 332.98  &   -24.03 &     12.62 &  7.0$\pm$1, 7.0, 7.5      \\
2016-06-13T11h31 & green triangle  & 336.93   &  -11.69  &    2.11 &  13.5$\pm$2, 13.5, 14.5       \\
2016-06-13T11h31 & magenta square  & 329.74 &    -23.29 &      5.34 & --       \\ \hline

2016-07-02T15h22 & circle    & 340.53  &  -15.47   &   18.87 & 0, DT     \\
2016-07-02T15h22 & red star & 346.15  &   -16.19  &   -3.87 &  in shadows        \\
2016-07-02T15h22 & blue star  &   335.31 &    -11.83 &    -4.08 &  26.5,27,29$\pm$3 \\
2016-07-02T15h22 & green triangle  &  344.35  &    -13.32  &    18.99 & --\\
2016-07-02T15h22 & magenta square  &  334.44  &   -10.73  &    -3.12 &  64,65.5$\pm$6,69.5 \\\hline

        \end{tabular}
\end{center}
\tablefoot{Symbols refer to the ROIs represented in Fig.~\ref{spotROI}. The water-ice abundance was estimated using a linear mixing model of water ice and cometary dark terrain, and reported for three different grain size (abundances with errors refer to the best fit model).  Errors in the spectral slope are on the order of 0.5\%/100nm.}
 \end{table*}
\begin{figure}[t]
\centering
\includegraphics[width=0.49\textwidth]{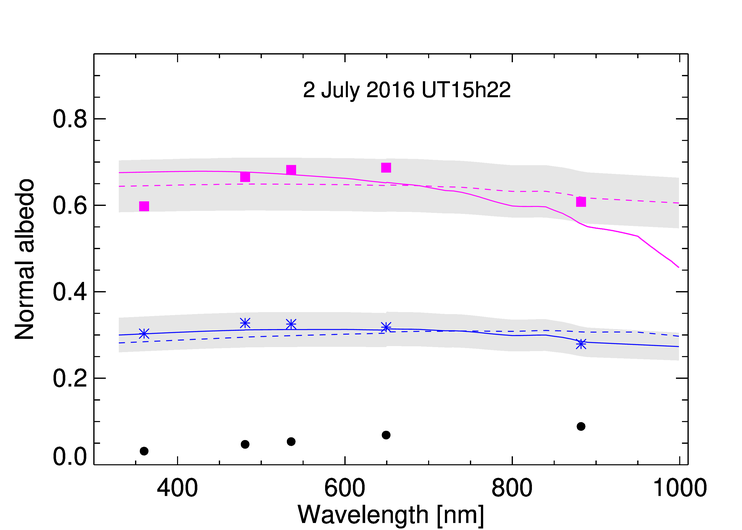}
\caption{Normal albedo of the bright spots (indicated by the magenta squares and the blue asterisks, see Fig.~\ref{spotROI}) observed on 2 July 2016 UT15h22. The symbols correspond to the bright spots shown on the right panel of Fig.~\ref{spotROI}. The black circles represent the mean spectrum of the comet from a region close to the bright patches. Continuous and dashed lines show the best fit spectral models to the bright patches (in gray the associated uncertainty), produced by the linear mixture of the comet dark terrain enriched with 29$\pm$3\% of water ice with 1000 $\mu$m grain size (continuous line) for the ROI indicated by the blue asterisk, and with 65.5$\pm$6.0\% of water ice with 100 $\mu$m grain size (dashed line)  for that represented by the magenta square. Models including water ice of medium and large grain size are represented for comparison: dashed lines indicate models with water ice of 100 $\mu$m grain size, and continuous lines those with 1000 $\mu$m grain size.} 
\label{icemodel}
\end{figure}
For the 13 June 2016 data, the ROI indicated by the blue asterisk (Fig.~\ref{spotROI}) is a typical example of a relatively blue terrain just emerging from shadows, where the spectral slope is lower ($\sim$ 12 (\%/100 nm)) than that of the average dark terrain, but far from the null value. We estimate a local enrichment of $\sim$ 7\% in water ice in this ROI, and the best fit is given by small water ice particles. The ROIs indicated by the red star and the green triangle have a flat spectrophotometric behavior indicating some exposure of water ice. They are close to shadows and very tiny, their bright surface being only of $\sim$ 1m$^2$, that is smaller than the box (2.3 m$^2$) used to integrate the signal. Therefore, the water-ice estimation (around 14\%  for the ROI indicated by the green triangle in Table~\ref{tab_model}) should be taken with caution because the integrated signal is partially affected by the shadowed regions. We show these ROIs as example of tiny BS seen on Wosret. \\ 
Of greater interest are the BS observed on 2 July 2016 at a spatial resolution that is about the double than in June 2016 (Fig.~\ref{spotROI}, right image). These BS are indicated by the magenta square and the blue asterisk, they also have a small size ($\sim$ 1.2 and 1.5 m$^2$, respectively), but they are larger than the box where the signal was integrated and they are extremely bright, by a factor of 12 and 5, respectively,  than the comet dark terrain. Their spectrophotometry is peculiar, showing a moderate negative slope in the 535-882 nm range. These ROIs are inside shadows and represent the top of morphological features illuminated only in these areas. Our best-fit compositional models for these two ROIs are shown in Fig.~\ref{icemodel}. The BS represented by the blue asterisk is best fit by a model including 29$\pm$3\% water ice with large grain size (1000 $\mu$m), which better mimics the decrease of reflectance at 880 nm than the model with smaller grain size. For the ROI indicated by the magenta square the estimated normal reflectance is very high, about 60-70\%, as it is the associated water-ice abundance: 65.5$\pm$6.0\% for water ice with a grain size of 100 $\mu$m.  However, these models have a higher chi-squared value than those reproducing the blue asterisk ROIs normal albedo and  do not fully match the observed spectrophotometry of the bright spot. In this case, we favor the model including medium size water ice grains (100 $\mu$m) because the one with larger grains produces a strong decrease in reflectance in the NIR range, which is not observed in the data. \\
Although the models are not optimal, we have clear indications of a local water ice abundance beyond 50\%, which is among the highest reported in the literature for bright spots on comet 67P, but in very small areas. Other tiny bright spots have been observed in Wosret in the area surrounding the Abydos landing site, including part of the Wosret, Bastet, and Hatmehit regions, which were recently investigated by Hoang et al (2020). They found a number of bright spots with size ranging from  0.1 m$^2$ to 27 m$^2$, having a distinct flat spectrum compared to the comet average dark and red terrain, and with an estimated local ice abundance reaching 50-80\% for 3 of them, thus indicating a  fresh exposure of volatiles. A high content of water ice ($\sim$ 46\%) was also determined for the boulder onto which Philae stamped two minutes during the second landing site located 30 m away from Abydos, unveiling the primordial water ice inside it (O'Rourke et al. 2020). \\
Other ROIs exposing fresh water ice with an estimated ice content higher than $\sim$ 25\%  were found in several areas in the Anhur, Bes, Khonsu, and Imhotep regions (Fornasier et al. 2016, 2017, 2019a, Deshapriya et al. 2016, Oklay et al. 2017, Hasselmann et al. 2019), as well as in the Aswan site (Pajola et al. 2017). It should be noted that these last bright and water-enriched areas are usually much more frequently observed and much larger than the ones detected in Wosret.
 
\begin{figure*}
\centering
\includegraphics[width=1.0\textwidth]{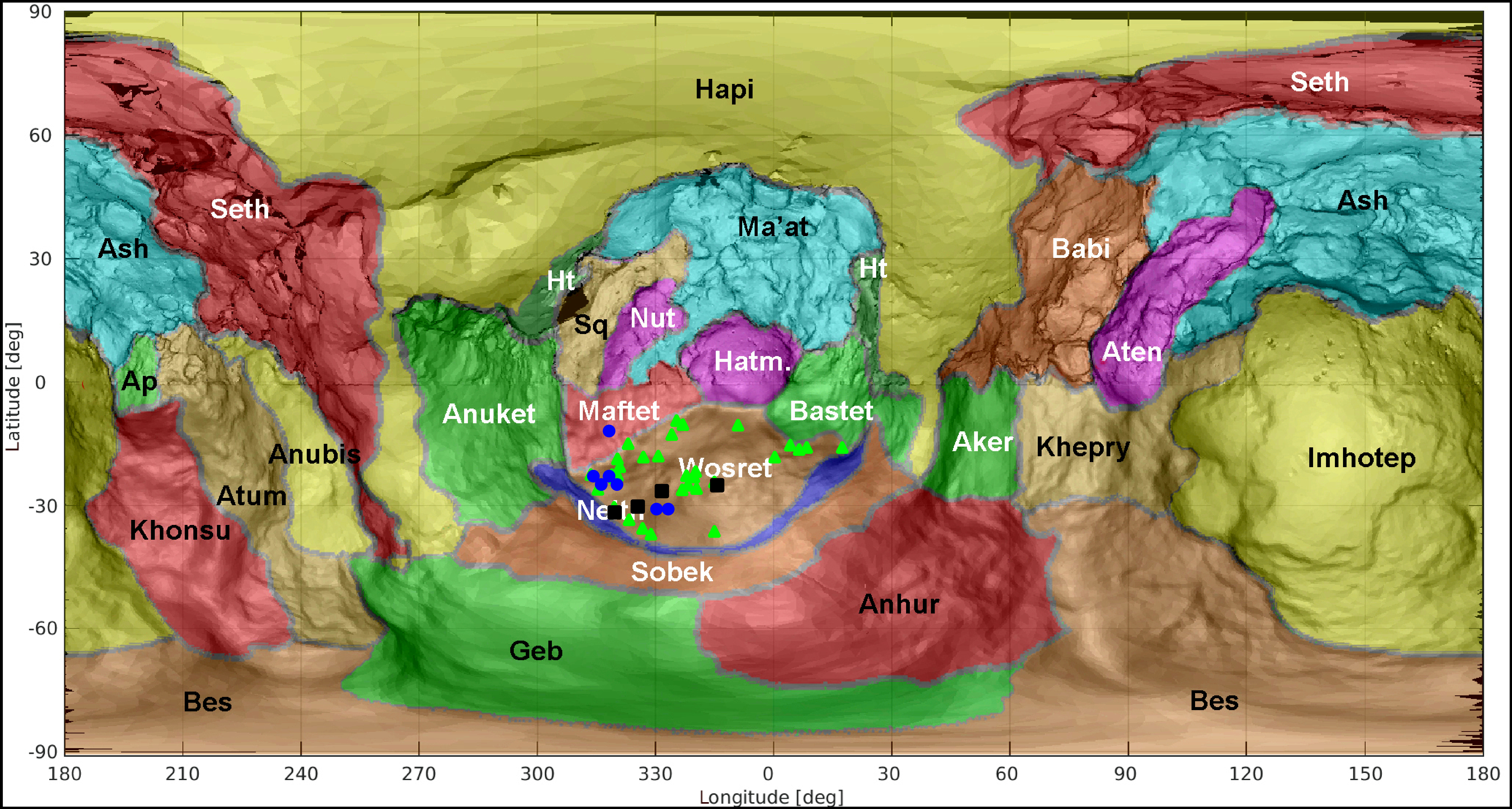}
\caption{Map of the comet with superposed the sources of jets identified in the Wosret region during the comet southern hemisphere summer. The location of some nearby jets are averaged for clarity, and some notable jets are represented with larger symbols. Blue circles represent jets reported in Vincent et al. (2016), green triangles those reported in Fornasier et al. (2019a), and black squares represent  cavities showing repeated activity in several data sets (Fornasier et al. 2019a).} 
\label{mapjets}
\end{figure*}
     
\section{Activity}

Wosret is also one of the most active regions: it has the highest estimated water production rate, which peaks at $\sim$ 10$^{28}$ molecules/s during the perihelion passage (Marshall et al. 2017), and it is the source of about 40 activity events, including one of the brightest outburst caught by the Rosetta observations (Vincent et al. 2016). In fact, 33 distinct source regions were reported by Fornasier et al. (2019b) and six more were identified by Vincent et al. (2016). Their position is shown in Figure~\ref{mapjets}, with black squares representing four cavities showing repeated activity close to the perihelion passage (see Fig. 18 in Fornasier et al. 2019b). These cavities were showing activity events in 16 to 60 distinct observations. The jets departing from these cavities are often relatively faint, sometimes with peculiar morphology. \\
On Wosret, the activity mechanism is triggered by changes in local insolation linked to the diurnal and seasonal cycles of water on the nucleus. In fact, most of the activity sources, notably the periodically active cavities, are in or close to shadowed areas, as similarly observed for the majority of active sources of comet 67P during the perihelion passage (Fornasier et al. 2019b). In areas casting shadows, subsurface water ice may easily recondensed during the cometary night, and it may survive at the cometary surface as frost or ice until the area is sufficiently illuminated and heated by the Sun to produce its sublimation. \\
Fornasier et al. (2019b) reported the presence of tiny water ice patches in one of the active cavities, while for the others, unfortunately, no observations at high spatial resolution are  available. Some images caught however few areas enriched in volatiles, as discussed in the previous section. An interesting example is reported in Fig.~\ref{bluepatch}, which displays the southern part of Wosret at high resolution (310$< lon < 330 ^o$, -30 $, lat < -20 ^o$), including a blue patch located at latitude -21$^o$ and longitude 318$^o$, and few tiny bright spots. This patch has a moderate spectral slope (12.7 \%/(100 nm)) about 50\% lower than that of the average dark terrain (18.1 \%/(100 nm)), and it is located close to the source region of a summer jet identified by Vincent et al. (2016, jet \#11). Similar processes associated with the condensation and sublimation cycle of volatiles are expected to be ongoing in the other cavities of Wosret that show periodic activity.

\section{Discussion and conclusions}

\begin{table*}
         \begin{center} 
\small{
         \caption{Summary of the observed behavior and physical properties for the small and big lobe of comet 67P.}
         \label{differences}
        \begin{tabular}{|l|l|l|l| } \hline
    & {\bf Small lobe (Wosret)} & {\bf Big lobe} & {\bf References } \\ \hline \hline
incoming solar flux & $\sim$ 550 W m$^{-2}$  &  $\sim$ 550 W m$^{-2}$ (Anhur and Khonsu) & Marshall et al. 2017 \\ \hline
morphology          & Consolidated material that appears      &  Consolidated material with significant   & Thomas et al. 2018 \\ 
                    & highly fractured with occasional pits   &  intermediate scale roughness (Anhur)    & \\ \hline
exposed water ice    & in a few and small bright spots ($\sim m^2$) & in many bright spots and  & This work, \\ 
                     &                                              & some in big patches (1500 m$^2$) & Fornasier et al. 2016, 2019a  \\ \hline
relatively blue area            & in very spatially limited area,    &  in extended area, periodic & This work, Hasselmann et al. 2018, \\
enriched in frost/ice              & periodic, not frequently observed  &  often observed             & Fornasier et al. 2016, 2017, 2019 \\ \hline
average goosebump         & the largest in 67P: 4.7$\pm$1.5 m           & 2.2-3.2 m (Seth, Imhothep,  & This work, Sierks et al. 2015,  \\ 
diameter                                   &                                             &    Anubis, Atum)        &             Davidsson et al. 2016 \\ \hline
level of activity                  & very high  & very high  & This work, Fornasier et al. 2019b, \\
                                   &                                                &          &  Hasselmann et al. 2019\\ \hline
surface morphology            & few and not so important:                       & many and important : &  This work, \\
changes                                   & a relatively small cavity,   &  local dust removal up to 14m in depth,   &  Fornasier et al. 2017, 2019a, \\
                                   & local dust removal ($\sim$ 1m depth)  & new relatively big scarps and cavities, & Hasselmann et al. 2019 \\
                                   &    revealing a cluster of outcrops       & big vanishing structures, boulders   & \\
                                   &                                                                      & displacements\&fragmentation & \\ \hline
surface mass loss                  & $\sim$ 1.2$\times 10^6$ kg                                           &  $> 50 \times$10$^6$ kg in Anhur & This work, Fornasier et al. 2019a\\
     &                    &     $\sim 2\times$10$^8$ kg in Khonsu & Hasselmann et al. 2019   \\ \hline
        \end{tabular}
}
\end{center}
\tablefoot{These properties are derived from the investigation of southern hemisphere regions submitted to similar high heating at perihelion. For the small lobe, we consider the results on Wosret presented in this work, while for the big lobe, we refer to results from the literature mainly focused on the Anhur and Khonsu regions.}
 \end{table*}

Among the 26 regions of comet
67P, Wosret is shown to have unique geomorphological features. It is one of the regions receiving the highest solar flux, that is, between 400 and 600 W/m$^2$ at perihelion (Marshall et al. 2017). This high insolation is responsible for both of the high erosion and activity level in Wosret. The high erosion level produces the flattened-out aspect, the pervasive fracturing observed in the consolidated areas in its southern part, and, together with the activity, the overall lack of dust deposits compared to other regions (El-Maarry et al. 2016). Consequently, Wosret exposes part of the inner layers of the small lobe (Penasa et al. 2017). 

The final landing site of Philae, Abydos, is located in the northern part of Wosret, close to the boundary with the Hatmehit depression and Bastet region (Fig.~\ref{wosret}). Its spectrophotometric properties have been studied by Hoang et al. (2020), who investigated an area of about 5$\times$5$^{o}$ in longitude and latitude surrounding the Philae final landing site.  This terrain is as dark as the overall nucleus albedo and it shows relatively red colors, with the presence of some localized bright spots of 0.1 m$^2$ to 27 m$^2$ size rich in water ice. They also identified few possible morphological changes in this area with an estimated  total mass loss of 4.7-7.0$\times$10$^5$ kg.\\
Globally, the surface properties of Abydos are nearly indistinguishable from those of Wosret, thus, the results derived from the Philae in situ measurements may be considered representative of the whole Wosret region.  The ubiquitous fractures observed in Wosret should extend also at sub-meter scale, as derived for the surrounding of Philae landing site (Bibring et al. 2015, Poulet et al. 2016). The fact that the lander instrument detected no dust impacts  (Kr\"uger et al. 2015) may support the overall paucity of dust in Wosret, even if this could also be related to shielding effects by the cliffs close to Philae combined with the relatively low level of activity in November 2014. Regarding the compressive strength of Abydos, very different numbers ranging from a few tenths of Pa to 2 MPa were reported in the literature (Spohn et al. 2015,  Biele et al. 2015, Knapmeyer et al. 2018, Heinisch et al. 2019), although the measurements derived from the penetrator sensors, which give the higher strengths values, should be taken with caution because  they are affected by deployment uncertainties. 

A recent reconstruction of the final Philae trajectory allowed for the identification of an additional landing site touched during about 2 minutes before Philae stopped in Abydos. This side is located 30 m apart from Abydos and it is also located in Wosret (O'Rourke et al. 2020). During this landing, Philae collided with a boulder, producing some morphological changes, lifting up the dust covering the surface and uncovering primitive buried water ice in a 3.5 m$^2$ size spot, six to eight times brighter than the average comet. This confirms previous findings showing that the water ice is abundant beneath the dark dust layer and drives the observed seasonal and diurnal color variations (Fornasier et al. 2016). In fact, cliff collapses observed in Seth and Anhur regions also exposed underneath layers that are rich in  water ice (Pajola et al. 2017, Fornasier et al. 2019a). \\
O'Rourke et al. (2020) were also able to measure the local porosity of the boulder into which Philae stamped, finding high values (75\%) consistent with those determined previously for the overall nucleus (Kofman et al. 2015, Herique et al. 2019, P\"atzold et al. 2016). They also found an extremely low ($<$12 Pascals) value for the compressive strength, indicating that the mixture of dust and ice is extremely soft (O'Rourke et al. 2020) even inside the boulders.

As shown in the previous section, Wosret spectrophotometric properties are similar to those observed in other regions of the 67P nucleus. Here, we report the presence of local bright spots and relatively blue areas, located close to shadows and associated with the diurnal cycle of water ice. These last areas are however very small and less frequent than the water ice enriched zones observed in Anhur, a region in the big lobe that is subject to the same strong thermal heating during the cometary summer and that is also highly active (Fornasier et al. 2017, 2019a). \\
Even if no appreciable differences between the two lobes have been reported in the literature on the global surface composition (Capaccioni et al. 2014, Fornasier et al. 2015), and on the deuterium to hydrogen ($D/H$) ratio (Schroeder et al. 2019), our study on Wosret highlights some differences in the physical and mechanical properties among the two lobes, which are summarized in Table~\ref{differences}:\\
\begin{enumerate}
\item First, the water-ice enriched regions directly exposed at the surface of Wosret are less frequent and smaller in size than those observed in Anhur, where water-ice rich areas of sizes ranging from a few to 1600 m$^2$ were observed (Fornasier et al. 2016). 
\item We report the presence of spectrally bluer area enriched in frost located close to shadows and related to the diurnal cycle of water. However, these areas are less frequently observed in Wosret  than in other regions. For example, frost is repeatedly observed in Anhur inside shadowed regions (Fornasier et al. 2019b).
\item The comparison of Wosret surface evolution with that of southern hemisphere regions such as Anhur and Khonsu, which experience the same high incoming solar flux than Wosret at perihelion (see Fig. 6 of Marshall et al. 2017), indicates important differences. In fact, all these regions are  highly active, but the activity results in different surface re-shaping. Indeed important morphological changes are observed in Anhur: the formation of two new scarps exposing the subsurface water ice layer; the local removal of the dust coating up to $\sim$ 14$\pm$2 m in depth within a canyon-like structure;  vanishing structures of several tenth of meters in length; and boulder displacement and fragmentation, for a total mass loss estimation higher than 50 $\times$10$^6$ kg (Fornasier et al. 2017, 2019a). For Khonsu, another southern hemisphere region of the big lobe, Hasselmann et al. (2019) reported a number of morphological changes (boulder displacements, cavities formation, dust bank removal), producing an estimated mass loss of $\sim 2\times$10$^8$ kg. Conversely, no major morphological changes are observed in Wosret, except for the formation of the new cavity, and the dust coating removal tentatively estimated in $\sim$ 1 m depth, locally. This difference cannot be related to observational biases because these regions have been observed under similar spatial resolution and illumination conditions pre- and post-perihelion. 
\item Polygonal block in Wosret are on average two times larger than the goosebumps (or clods) observed in different regions on the big lobe. These structures  have been interpreted as being representative of the original cometesimals forming, by aggregation, cometary nuclei, or as the result of fracturing processes produced by the seasonal and diurnal thermal gradients in the material (Sierks et al. 2015, El-Maarry et al. 2015, Auger et al. 2016). 
\end{enumerate}

El-Maarry et al. (2016), based on the morphological analysis of the 67P regions, deduced that the material composing the two lobes has different physical and mechanical properties. We can support this conclusion by the fact that the clods-goosebumps features in Wosret are larger than those observed in the big lobe.  Moreover, the limited morphological changes observed in Wosret as compared to Anhur, shown in Table~\ref{differences} (while both regions were among the most active ones and the most heated by the Sun during the perihelion passage), indicate that the material on Wosret should be less fragile and more consolidated than the material present on the big lobe southern regions. Furthermore, the low number of exposed volatiles at the surface of Wosret, as compared to Anhur, might indicate that the small lobe has a lower volatile content, at least in its top layers, than the big lobe. \\

All this evidence support the hypothesis formulated by Massironi et al. (2015) based on the analysis of the layering of the two lobes stating that comet 67P is composed of two distinct bodies that merged during a low-velocity collision in the early Solar System.



\begin{acknowledgements}
OSIRIS was built by a consortium led by the Max-Planck-Institut f\"ur Sonnensystemforschung, Goettingen, Germany, in collaboration with CISAS, University of Padova, Italy, the Labo
ratoire d'Astrophysique de Marseille, France, the Instituto de Astrof\'isica de Andalucia,CSIC, Granada, Spain, the Scientific Support Office of the European Space Agency, Noordwijk, The Netherlands, the Instituto Nacional de T\'ecnica Aeroespacial, Madrid, Spain, the Universidad Polit\'echnica de Madrid, Spain, the Department of Physics and Astronomy of Uppsala University, Sweden, and the Institut f\"ur Datentechnik und Kommunikationsnetze der Technischen Universitat Braunschweig, Germany. \\
The support of the national funding agencies of Germany (DLR), France (CNES), Italy (ASI), Spain (MEC), Sweden (SNSB), and the ESA Technical Directorate is gratefully acknowledged.
 We thank the Rosetta Science Ground Segment at ESAC, the Rosetta Mission Operations Centre at ESOC and the Rosetta Project at ESTEC for their outstanding work enabling the science return of the Rosetta Mission. We acknowledge the financial support from the France Agence Nationale de la Recherche (programme Classy, ANR-17-CE31-0004). 

\end{acknowledgements}

%
%

%


%

%


\end{document}